\documentclass[useAMS,usenatbib]{mn2e}

\usepackage{graphicx}
\usepackage{txfonts}

\def\ion#1#2{{\rm #1}\,{\sc #2}}

\title[The high metallicities of sub-DLAs]
{Revisiting the origin of the high metallicities of sub-damped Lyman-alpha 
systems\thanks{Based on observations collected at the European Southern 
Observatory, Chile, under programmes ID No. 078.A--0003(A) and 080.A--0014(B).}}

\author[Miroslava Dessauges-Zavadsky, Sara L. Ellison, Michael T. Murphy]
{Miroslava Dessauges-Zavadsky$^{1}$\thanks{E-mail: 
miroslava.dessauges@unige.ch}, Sara L. Ellison$^{2}$, Michael T. Murphy$^{3}$\\
$^{1}$Observatoire de Gen\`eve, Universit\'e de Gen\`eve, 51 Ch. des 
Maillettes, 1290 Sauverny, Switzerland\\
$^{2}$Department of Physics and Astronomy, University of Victoria, Victoria, 
B.C., V8P 1A1, Canada\\
$^{3}$Centre for Astrophysics and Supercomputing, Swinburne University of 
Technology, Mail H39, PO Box 218, Victoria 3122, Australia}

\begin{document}


\pagerange{\pageref{firstpage}--\pageref{lastpage}} \pubyear{2002}

\maketitle

\label{firstpage}


\begin{abstract}
Sub-damped Lyman-alpha systems (sub-DLAs) have previously been found to exhibit 
a steeper metallicity evolution than the classical damped Lyman-alpha systems 
(DLAs), evolving to close to solar metallicity by $z\sim 1$. From new 
high-resolution spectra of 17 sub-DLAs we have increased the number of 
measurements of [Fe/H] at $z<1.7$ by 25\% and compiled the most complete 
literature sample of sub-DLA and DLA abundances to date. We find that sub-DLAs 
are indeed significantly more metal-rich than DLAs, but only at $z<1.7$; the
metallicity distributions of sub-DLAs and DLAs at $z>1.7$ are statistically
consistent. We also present the first evidence that sub-DLAs follow a velocity
width~--~metallicity correlation over the same velocity range as DLAs, but the 
relation is offset to higher metallicities than the DLA relation. On the basis 
of these results, we revisit the previous explanation that the systematically 
higher metallicities observed in sub-DLAs are indicative of higher host galaxy 
masses. We discuss the various problems that this interpretation encounters and 
conclude that in general sub-DLAs are not uniquely synonymous with massive 
galaxies. We rule out physically related sources of bias (dust, environment, 
ionization effects) and examine systematics associated with the selection and 
analysis of low-redshift sub-DLAs. We propose that the high metallicities of 
sub-DLAs at $z<1.7$ that drives an apparently steep evolution may be due to the 
selection of most low-redshift sub-DLAs based on their high \ion{Mg}{ii} 
equivalent widths. 
\end{abstract}



\section{Introduction}\label{intro}

Quasar absorption line (QAL) systems are very valuable tools for the census of 
metals from the present epoch all the way back to the farthest detectable 
quasars (QSOs). Much attention has recently been directed towards the 
sub-damped Lyman-alpha systems (sub-DLAs), QAL systems with \ion{H}{i} column 
densities in the range $10^{19} < N$(\ion{H}{i}) $< 2\times 10^{20}$ cm$^{-2}$. 
The sub-DLAs outnumber the higher \ion{H}{i} column density damped Lyman-alpha 
systems (DLAs) by factors of $4-8$ and may therefore contribute significantly 
to the cosmological metals budget. Two major surveys for high-redshift sub-DLAs 
have been undertaken \citep{DZ03,peroux05}, and recently much effort has been 
dedicated to building up a sub-DLA sample at $z<1.7$ \citep{meiring06,meiring07,
meiring08,meiring09,peroux06a,peroux08}. These surveys have yielded abundance 
measurements ([Fe/H] or [Zn/H]) for 30 and 35 sub-DLAs at redshifts above and 
below $z=1.7$, respectively.

\citet{peroux03} were the first to point out a stronger metallicity evolution 
with time in sub-DLAs than DLAs. This has been further supported by the 
enlarged sample of sub-DLAs and has led \citet{kulkarni07} to conclude: (1)~The 
mean metallicity of sub-DLAs is higher than that of DLAs and reaches a 
near-solar level at $z<1$. Several sub-DLAs at low redshifts even exhibit solar 
or supersolar abundances \citep{peroux06b,prochaska06}, while no DLA is known 
so far with such a high metallicity. (2)~The contribution of sub-DLAs to the 
total metal budget may be several times greater than that of DLAs at $z<1$. 
Nevertheless, sub-DLAs seem to contribute no more than a few percent of the 
total amount of metals in the Universe at $z\sim 2.5$ \citep{peroux07}. 

The difference in metallicity between sub-DLAs and DLAs has led \citet{khare07} 
to suggest that sub-DLAs and DLAs may be associated with two distinct 
populations of galaxies. Indeed, since high metallicities are generally 
associated with more massive galaxies according to the mass~--~metallicity 
relation \citep{tremonti04,savaglio05,erb06,maiolino08}, \citet{khare07} 
concluded that sub-DLAs may arise in massive spiral/elliptical galaxies and 
DLAs in low mass ($<10^9~M_{\odot}$) galaxies.

\begin{table*}
\centering
\begin{minipage}{160mm}
\caption{New sub-DLA systems.}
\label{observations}
\begin{tabular}{@{}llccclccc@{}}
\hline
QSO name & Other name & $V$   & $z_{\rm QSO}$ & $z_{\rm sub-DLA}$ & Instrument & $\log N$(\ion{H}{i}) & [Fe/H] & [Zn/H] \\
         & (J2000)    & [mag] &               &                   &            & [cm$^{-2}$]          &        &        \\
\hline
QSO B0009$-$016  & J001210.9$-$012208 & 17.6 & 1.998 & 1.3861 & UVES/VLT     & $20.26\pm 0.02$ & $-1.39\pm 0.04$ & $<-2.02$ \\
QSO J0021+0043   & J002133.3+004300   & 17.7 & 1.245 & 0.5203 & UVES/VLT     & $19.54\pm 0.03$ & $-1.82\pm 0.05$ & ...      \\
...              & ...                & ...  & ...   & 0.9424 & ...          & $19.38\pm 0.13$ & $-0.21\pm 0.14$ & $<-0.41$ \\
QSO J0157$-$0048 & J015733.8$-$004824 & 17.9 & 1.548 & 1.4157 & UVES/VLT     & $19.90\pm 0.07$ & $-0.78\pm 0.08$ & $-0.43\pm 0.11$ \\
QSO B0216+08     & J021857.3+081728   & 18.1 & 2.996 & 1.7687 & UVES archive & $20.20\pm 0.10$ & $-1.17\pm 0.10$ & $-0.81\pm 0.16$ \\
QSO B0352$-$275  & J035405.6$-$272420 & 17.9 & 2.823 & 1.4054 & UVES/VLT     & $20.18\pm 0.15$ & $-0.53\pm 0.15$ & $+0.05\pm 0.15$ \\
QSO B0424$-$131  & J042707.3$-$130254 & 17.5 & 2.166 & 1.4080 & UVES/VLT     & $19.04\pm 0.04$ & $-1.04\pm 0.04$ & $<-0.84$ \\
QSO J1009$-$0026 & J100930.5$-$002618 & 17.4 & 1.244 & 0.8428 & UVES/VLT     & $20.20\pm 0.06$ & $-1.17\pm 0.06$ & $<-1.41$ \\
...              & ...                & ...  & ...   & 0.8865 & ...          & $19.48\pm 0.05$ & $-0.56\pm 0.09$ & $+0.24\pm 0.15$ \\
QSO J1028$-$0100 & J102837.1$-$010028 & 18.2 & 1.531 & 0.6321 & UVES/VLT     & $19.95\pm 0.07$ & $-0.34\pm 0.08$ & $<-0.20$ \\
...              & ...                & ...  & ...   & 0.7089 & ...          & $20.04\pm 0.06$ & $-0.39\pm 0.07$ & $<-0.18$ \\
QSO B1037$-$270  & J103921.9$-$271916 & 17.4 & 2.193 & 2.1395 & UVES archive & $19.60\pm 0.10$ & $-0.35\pm 0.10$ & $-0.05\pm 0.10$ \\
QSO J1054$-$0020 & J105441.0$-$002048 & 18.3 & 1.021 & 0.9513 & UVES/VLT     & $19.28\pm 0.02$ & $-1.02\pm 0.02$ & $<-0.66$ \\
QSO B1327$-$206  & J133007.8$-$205617 & 17.0 & 1.165 & 0.8514 & UVES/VLT     & $19.40\pm 0.02$ & $-0.95\pm 0.04$ & $<-0.46$ \\
QSO J1525+0026   & J152510.6+002633   & 17.0 & 0.801 & 0.5674 & HIRES/Keck   & $19.78\pm 0.08$ & $-1.04\pm 0.10$ & ...      \\
QSO B2048+196    & J205112.7+195007   & 18.5 & 2.367 & 1.1161 & HIRES/Keck   & $20.00\pm 0.15$ & $>-0.23$        & $+0.33\pm 0.09$ \\
QSO J2352$-$0028 & J235253.5$-$002852 & 18.2 & 1.628 & 0.8730 & HIRES/Keck   & $19.18\pm 0.09$ & $-1.13\pm 0.11$ & $<+0.22$ \\
...              & ...                & ...  & ...   & 1.0318 & ...          & $19.81\pm 0.13$ & $-0.30\pm 0.14$ & $<-0.50$ \\
...              & ...                & ...  & ...   & 1.2468 & ...          & $19.60\pm 0.24$ & $-0.77\pm 0.24$ & $<-1.13$ \\ 
\hline
\multicolumn{9}{l}{\textit{Note}: Values reported as lower limits refer to 
saturated lines and values reported as upper limits correspond to $4\,\sigma$ 
non-detections.}
\end{tabular}
\end{minipage}
\end{table*}

In this Letter, we form the largest sample of low-redshift sub-DLAs with 
abundance determinations, and we present evidence against the hypothesis that 
sub-DLAs are in general associated with massive galaxies.


\section{Samples, observations, column densities}\label{sample}

Our sample of sub-DLAs consists of 17 systems at $z<1.7$ chosen from the 
compilation of \citet{rao06}\footnote{Since the submission of this Letter, 
\citet{meiring09} have obtained data for a smaller sub-DLA sample with which we 
have 4 QSOs in common.}. We have obtained high resolution ($FWHM\simeq 7-8$ 
km~s$^{-1}$), high signal-to-noise ratio (S/N $>10$ per pixel from 3300 to 6800 
\AA) spectra with the Ultraviolet-Visual Echelle Spectrograph, UVES, on the VLT 
Kueyen ESO telescope at Cerro Paranal, Chile and the High Resolution Echelle 
Spectrometer, HIRES, on the Keck~I telescope at Mauna Kea, Hawaii. Our sample 
is supplemented with 2 additional sub-DLAs at $z>1.7$ toward QSOs B0216+08 and
B1037$-$270 retrieved from the UVES ESO/ST-ECF Science Archive 
Facility\footnote{Programmes ID No. 65.P--0038(A), 69.B--0108(A), 
70.A--0017(A), 71.B--0136(A), 072.A--0346(A), and 073.B--0787(A).}. We reduced 
the UVES and HIRES echelle spectra using, respectively, the publically 
available UVES\_popler 
pipelines\footnote{http://astronomy.swin.edu.au/$\sim$mmurphy/UVES\_popler} 
and the xidl HIRED redux 
pipeline\footnote{http://www.ucolick.org/$\sim$xavier/HIRedux/index.html}. 
The ionic column densities were measured using the Voigt profile fitting 
technique and the $\chi^2$ minimization routine \texttt{FITLYMAN} in 
\texttt{MIDAS}. The \ion{H}{i} column densities were taken from \citet{rao06} 
obtained from Hubble Space Telescope (HST) spectra, except for the sub-DLA
toward QSO~B2048+196 for which we considered the revised $N$(\ion{H}{i}) value
by \citet{meiring09}. For the 2 sub-DLAs at $z>1.7$ toward QSOs B0216+08 and 
B1037$-$270, the $N$(\ion{H}{i}) were obtained from UVES archive spectra. A 
complete description of the data acquisition, reduction, and full abundance 
analysis will be presented in a forthcoming paper (Dessauges-Zavadsky et~al., 
in preparation).


\begin{figure*}
\centering
\includegraphics[width=16cm,clip]{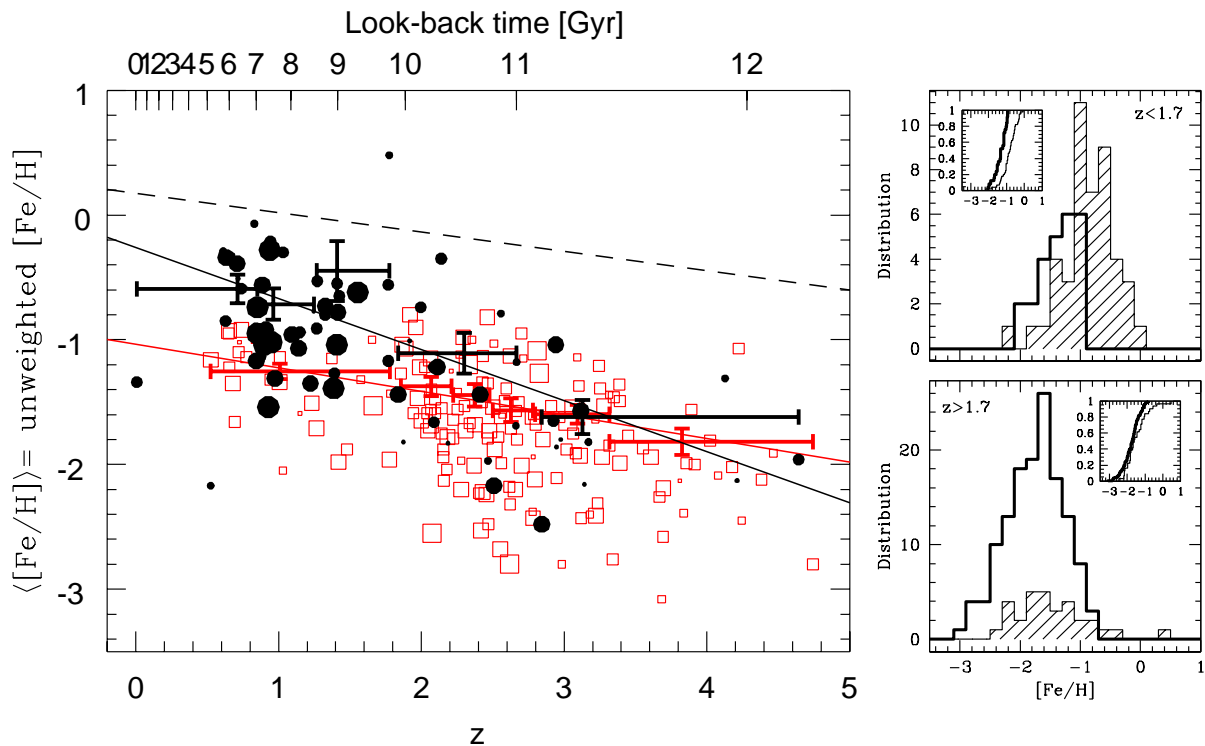}
\caption{[Fe/H] metallicities of 76 sub-DLAs (black filled circles) and 161 
DLAs (red open squares) plotted as a function of redshift, with the area of 
each symbol inversely scaled to the $1\,\sigma$ error on the abundance. The 
large black and red crosses correspond, respectively, to the unweighted mean 
metallicities of sub-DLAs and DLAs. In each bin, the mean is plotted at the 
median redshift, with the horizontal bars indicating the redshift range spanned 
by the systems in that bin. The least-square fits to the unweighted means are 
shown by the black solid line for sub-DLAs and the red solid line for DLAs. For 
comparison, the dashed line gives the metallicity evolution of the star-forming 
galaxies \citep{li08}. The two plots on the right-hand side show the [Fe/H] 
distributions of sub-DLAs (shaded histograms) and DLAs (open histograms) with 
their respective cumulative functions in two separate redshift intervals, 
$z<1.7$ (upper panel) and $z>1.7$ (lower panel).}
\label{metallicity-evolution}
\end{figure*}

In Table~\ref{observations} we summarize our sub-DLA sample and the derived 
abundances of Fe and Zn. The sample includes 6 new sub-DLAs with close to solar 
metallicities ($>1/5$ solar) and 3 systems with super-solar metallacities. With 
these new data, we enlarge the sample of sub-DLAs at $z<1.7$ with metallicity 
measurements ([Fe/H] or [Zn/H]) by 25\%, bringing it to a total of 47 systems; 
32 systems at $z>1.7$ complete the sub-DLA sample (for references see 
Sect.~\ref{intro}). For comparison, we use our compilation of DLA metallicity 
measurements, the most complete to date, which accounts 27 systems at $z<1.7$ 
and 144 at $z>1.7$ (Dessauges-Zavadsky et~al., in preparation).


\section{Results}\label{results}

Zn is often favoured as a metallicity indicator in gas-phase studies, since it 
is nearly undepleted and traces Fe very closely in Galactic stars 
\citep[e.g.,][]{pettini99}. However, there are still very few [Zn/H] 
measurements available, particularly in sub-DLAs at high redshifts (6 [Zn/H]
measurements at $z>1.7$). Conversely, [Fe/H] abundances are available for the 
majority of absorbers. Although Fe is depleted onto dust grains, the 
distributions of [Zn/Fe] ratios in sub-DLAs and DLAs at $z<1.7$ are 
statistically consistent (Kolmogorov-Smirnov (KS) confidence level (c.l.) to 
reject the null-hypothesis $= 23$\%; see Dessauges-Zavadsky et~al., in 
preparation). This suggests that sub-DLAs and DLAs (at least at low redshifts) 
have, on average, similar differential dust depletion levels. Hence, we should 
still be able to compare the sub-DLA and DLA metallicities and evolution 
thereof differentially, even if the absolute values are affected by dust 
depletion.

Fig.~\ref{metallicity-evolution} shows the [Fe/H] measurements obtained in 
sub-DLAs and DLAs plotted as a function of redshift. The large black and red 
crosses represent, respectively, the unweighted mean metallicities, $\langle 
[{\rm Fe/H}]\rangle$, of sub-DLAs in 5 redshift bins containing 16 or 12 
systems each and of DLAs in 6 redshift bins containing 27 or 26 systems each. 
The $1\,\sigma$ errors of the means include sampling and measurement 
uncertainties and are calculated using the statistical techniques described in 
\citet{kulkarni02}. The unweighted mean metallicities as well as the \ion{H}{i} 
column density weighted mean metallicities, $[\langle {\rm Fe/H}\rangle] = 
\log [\sum_i 10^{{\rm [Fe/H]}_i} N$(\ion{H}{i})$_i / \sum_i N$(\ion{H}{i})$_i]$, 
derived in each redshift bin are tabulated in online material. The unweighted 
means refer to the average metallicity of galaxies, while the \ion{H}{i} 
weighted means refer to the mass-weighted metallicity of neutral gas. Since the 
bulk of $N$(\ion{H}{i}) is not in sub-DLAs, the unweighted means are the 
appropriate quantity to trace and compare the metallicity evolution of the two 
samples of absorbers.

Clearly, the mean [Fe/H] metallicity of sub-DLAs at low redshifts is 
substantially larger than that of DLAs by about 0.7~dex ($4\,\sigma$ 
significance level). At $z>1.7$ the mean metallicity of sub-DLAs drops by 
0.5~dex and reaches a value similar to the mean metallicity of DLAs. This 
reveals two major differences in the metallicity evolution of sub-DLAs and 
DLAs. First, \textit{the sub-DLAs show a more rapid metallicity evolution with 
time than the DLAs}. Performing a least-square fit to the unweighted mean 
[Fe/H] metallicity versus redshift data points, we find for sub-DLAs a slope 
$m = -0.41\pm 0.07$ dex/$\Delta z$ and a zero point $b = -0.26\pm 0.13$ dex, 
while the slope for DLAs is less than half this value, $m = -0.19\pm 0.04$ 
dex/$\Delta z$, with a zero point $b = -1.03\pm 0.09$ dex. The best-fit slopes 
to the \ion{H}{i} weighted mean metallicities are identical within statistical 
uncertainties, both for sub-DLAs and DLAs, to the slopes derived for the 
unweighted mean metallicities. They underline the robustness of this result, 
previously claimed by \citet{peroux03,peroux07} and \citet{kulkarni07}. 

Second, \textit{the metallicity difference between sub-DLAs and DLAs is 
confined to low redshifts}. This is well illustrated in the two panels on the 
right-hand side of Fig.~\ref{metallicity-evolution} that show the [Fe/H] 
distributions of sub-DLAs and DLAs in two separate redshift intervals, $z<1.7$ 
and $z>1.7$. At $z<1.7$ sub-DLAs are clearly more metal-rich than DLAs, while 
at $z>1.7$ this large metallicity difference is no longer apparent (see the 
respective cumulative functions). The KS test shows that the null-hypothesis 
that the metallicities of the low-redshift sub-DLAs and DLAs are drawn from the 
same population can be rejected at 99.9992\% c.l. This c.l. falls to an 
inconclusive 90\% for the metallicity distributions of the high-redshift 
sub-DLAs and DLAs. This suggests that at high redshifts sub-DLAs have similar 
metallicities to DLAs, which diverge from that of DLAs at low redshifts only. 
With a mean metallicity [Zn/H] $=+0.18\pm 0.11$ in sub-DLAs and [Zn/H] 
$= -0.66\pm 0.09$ in DLAs at $z< 1.7$, the [Zn/H] measurements further confirm 
the bias of low-redshift sub-DLAs towards higher metallicities relatively to 
DLAs.

\begin{figure}
\centering
\includegraphics[height=7.7cm,clip]{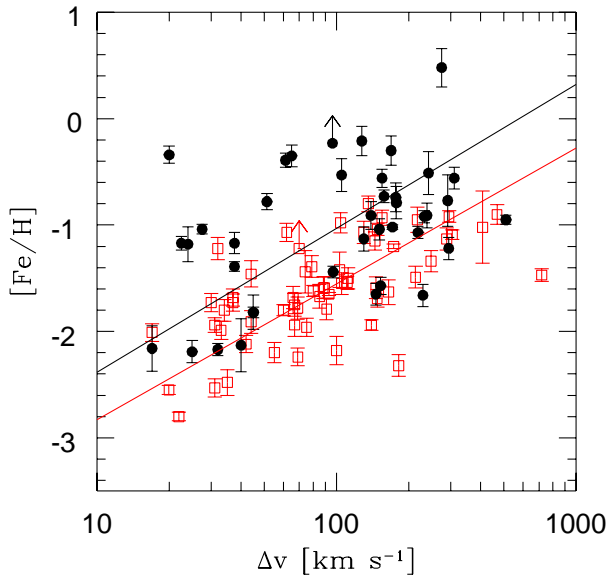}
\caption{[Fe/H] metallicities of 40 sub-DLAs (black filled circles) and 59 DLAs
(red open squares) plotted as a function of the velocity widths of their 
low-ionization line profiles displayed on a logarithmic scale. The linear 
least-square bisector fits are shown by the black solid line for sub-DLAs and 
the red solid line for DLAs.}
\label{velocity-metallicity}
\end{figure}

\citet{ledoux06} found a correlation between the velocity widths, $\Delta 
{\rm v}$, of lines of low-ionization species and metallicities, over two orders 
of magnitude in metallicity, for DLAs and a few high-$N$(\ion{H}{i}) sub-DLAs 
at $1.7<z<4.3$. Assuming that $\Delta {\rm v}$ is a measure of mass, it has 
been suggested that this velocity width~--~metallicity relation may be the 
consequence of an underlying mass~--~metallicity relation, consistent with the 
mass~--~metallicity relation observed for local and high-redshift galaxies (for 
references see Sect.~\ref{intro}). The mass~--~metallicity relation for DLAs 
has, in addition, been recently reproduced from simulations by 
\citet{pontzen08}.

Fig.~\ref{velocity-metallicity} shows the velocity widths and metallicities 
for sub-DLAs and DLAs of our sample and our literature compilation. The 
velocity widths of lines of low-ionization species were derived following the 
prescriptions of \citet{ledoux06}. With our enlarged sample of sub-DLAs, it can 
be seen that \textit{sub-DLAs also exhibit a correlation between the velocity 
widths and metallicities over the same range of $\Delta {\rm v}$ as DLAs} 
(Kendall's tau probability of 0.016\%), but with a slightly larger scatter than 
DLAs and offset to higher metallicities by $0.4-0.6$~dex at a given $\Delta 
{\rm v}$ (see the linear least-square bisector fits). The KS test shows that 
the null-hypothesis that the velocity widths of sub-DLAs and DLAs are drawn 
from the same population can be rejected at only 92\% c.l. Our results 
therefore contrast with the conclusions drawn from smaller sub-DLA samples: 
(i)~\citet{bouche08} argued that sub-DLAs have roughly constant metallicities 
as a function of the \ion{Mg}{ii} equivalent width; and (ii)~\citet{meiring07} 
obtained a velocity width~--~metallicity relation in which all the low-redshift 
sub-DLAs were confined to the upper right corner of 
Fig.~\ref{velocity-metallicity}, with high metallicities and larger velocity 
widths. Although there is still a tendency for sub-DLAs to have higher velocity 
widths (59\% of sub-DLAs have $\Delta {\rm v} > 100$ km~s$^{-1}$ compared to 
43\% of DLAs), Fig.~\ref{velocity-metallicity} clearly shows that sub-DLAs 
inhabit the same $\Delta {\rm v}$ range as DLAs. 


\section{Discussion and conclusions}\label{discussion}

With our enlarged sample of sub-DLAs, we conclude: (1)~sub-DLAs have a more 
rapid metallicity evolution with time than DLAs (in agreement with previous 
studies); (2)~sub-DLAs are significantly more metal-rich than DLAs, but only at 
$z<1.7$; and (3)~sub-DLAs follow a velocity width~--~metallicity correlation, 
although the relation is offset to higher metallicities than the DLA relation.

On the basis of the high metallicities observed in sub-DLAs and the existence 
of a mass~--~metallicity relation amongst galaxies at $0< z<3$, \citet{khare07} 
proposed that sub-DLAs arise in more massive galaxies than DLAs. With the new 
data presented in this Letter, this interpretation encounters a number of 
problems.

The \textit{first} problem is directly related to the main argument that led 
\citet{khare07} to link sub-DLAs to massive galaxies, namely the 
mass~--~metallicity relation and the velocity width~--~metallicity relation. If 
the velocity width of lines of low-ionization species in the velocity 
width~--~metallicity relation of \citet{ledoux06} is a proxy for mass and 
sub-DLAs are associated mostly with massive galaxies, we expect sub-DLAs to 
exhibit almost exclusively large velocity widths. As discussed above and 
illustrated in Fig.~\ref{velocity-metallicity}, sub-DLAs actually have 
velocity widths that span the full range of values of DLAs. 

\begin{figure}
\centering
\includegraphics[height=7.8cm,clip]{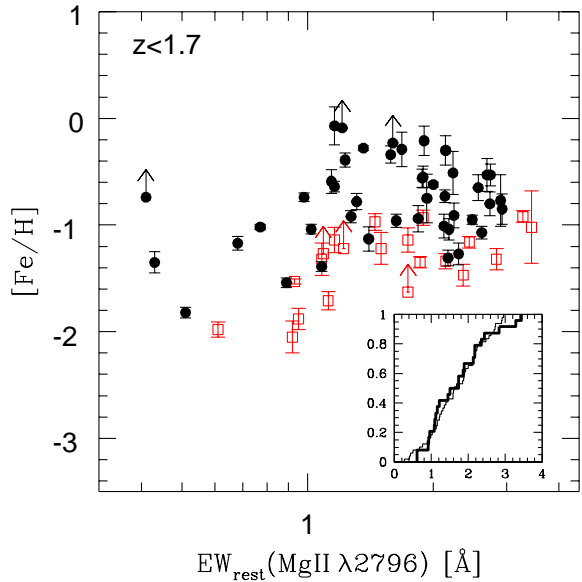}
\caption{[Fe/H] metallicities of 49 sub-DLAs (black filled circles) and 24 DLAs 
(open red squares) at $z<1.7$ plotted as a function of the rest-frame 
\ion{Mg}{ii}\,$\lambda$2796 equivalent widths displayed on a logarithmic scale.
The plot on the right-bottom corner shows the cumulative functions of the 
\ion{Mg}{ii}\,$\lambda$2796 EW distributions of sub-DLAs (thin line) and DLAs 
(thick line).}
\label{EW-metallicity}
\end{figure}

The \textit{second} difficulty with the identification of sub-DLAs with massive 
galaxies is to reconcile their metallicity evolution with the now popular 
scenario of `down-sizing'. In this paradigm, massive galaxies build up their 
stellar mass and metals at earlier epochs than lower mass galaxies and already
are relatively metal-rich by $z\sim 2$ \citep[e.g.,][]{panter08}. Contrary to
what is expected of massive galaxies, sub-DLAs are fairly metal-poor at high 
redshifts and only develop their metallicities at $z<1.7$ (see
Fig.~\ref{metallicity-evolution}). Moreover, the relative rates of sub-DLA and 
DLA metallicity evolution are inconsistent with a simple division in mass 
\citep[see e.g. Figure 3 of][]{panter08}. However, when splitting our full 
sample of sub-DLAs plus DLAs by their velocity width (and not their 
$N$(\ion{H}{i})), we observe at high redshifts, $z>2$, that systems with 
$\Delta {\rm v} > 100$ km~s$^{-1}$ have higher metallicities 
($\langle [{\rm Fe/H]}\rangle = -1.29\pm 0.07$) than systems with $\Delta 
{\rm v} < 100$ km~s$^{-1}$ ($\langle [{\rm Fe/H]}\rangle = -1.53\pm 0.18$). The 
KS test shows that the null-hypothesis that the metallicities of these two 
sub-samples of absorbers at $z>2$ are drawn from the same population can be 
rejected with confidence at 99.9991\% c.l. This result is in line with the 
expectations of both the down-sizing picture and the velocity width as a proxy 
for mass also supported by the recent simulations of \citet{pontzen08}. Neither 
sub-DLAs nor DLAs appear as the dominant feature of the sub-sample of absorbers 
at $z>2$ with $\Delta {\rm v} > 100$ km~s$^{-1}$ or the sub-sample of absorbers 
at $z>2$ with $\Delta {\rm v} < 100$ km~s$^{-1}$: the former (the latter) 
sub-sample is composed of 22\% (19\%) sub-DLAs and 78\% (81\%) DLAs. This 
suggests that the $N$(\ion{H}{i}) limit between sub-DLAs and DLAs is not simply 
due to a distinction in mass.

The \textit{third} objection to the massive galaxy hypothesis is the slope of 
the metallicity evolution of sub-DLAs which appears to be much steeper than 
that of luminous (massive) star-forming galaxies at $1<z<3$ \citep[e.g.,][and 
see the dashed line in Fig.~\ref{metallicity-evolution}]{li08}.

Finally, the \textit{fourth} inconsistency in connecting sub-DLAs with massive 
galaxies is that they outnumber DLAs by a factor of up to 8, whereas massive 
galaxies are far less common than low mass galaxies. Even when assuming that 
only the most metal-rich sub-DLAs arise in massive galaxies, at $z<1.7$ more 
than 70\% of sub-DLAs have [Zn/H] metallicities $> 2/5$ solar and hence lead to 
a too large proportion of massive galaxies. Although massive gas-rich galaxies 
may have a higher absorption cross-section, \citet{rosenberg03} have shown that 
in the local Universe the number density of DLAs is still dominated by 
sub-$M_{\star}$ galaxies.

So, how can the high metallicities of the low-redshift sub-DLAs be explained\,? 
It has often been suggested that dust in the higher column density DLAs leads 
to an observed distribution of metallicities that is biased towards low values. 
Indeed, if dust extinction is a strong function of the metal column density as 
observed in interstellar medium clouds \citep[e.g.,][]{vladilo04}, the 
metal-rich DLAs may be more affected by dust selection than metal-rich sub-DLAs 
\citep{vladilo05,meiring08,peroux08}. Such an effect could make DLAs appear 
less metal-rich than sub-DLAs. So far, however, there is no evidence at either 
high or low redshifts for a large number of high-extinction DLAs with elevated 
metallicities \citep[e.g.,][]{ellison01,murphy04,ellison05,akerman05,pontzen08,
pontzen09} that would be required to explain the $+0.7$~dex difference in 
metallicity between the low-redshift sub-DLAs and DLAs. Most recently, 
\citet{ellison09a} found that dust depletions in $0.6<z<1.7$ sub-DLAs and DLAs 
in the Complete Optical and Radio Absorption Line System (CORALS) survey are 
consistent with magnitude limited samples. 

Can large-scale physical differences account for the metallicity differences
between low-redshift sub-DLAs and DLAs\,? Recently, \citet{ellison08} have 
shown that for a given mass, galaxies are more metal-rich if they have smaller 
half-light radii or lower specific star formation rates. \citet{ellison09b} 
have also shown that galaxies in dense environments have slightly higher 
metallicities. However, the metallicity offsets for both effects are far too 
small ($\sim 0.05$~dex) to explain the differences between sub-DLAs and DLAs we
observe.

What about photoionization effects\,? The lower \ion{H}{i} column densities of 
sub-DLAs may imply that the neutral hydrogen is partially ionized, in which 
case the observed \ion{Fe}{ii} to \ion{H}{i} ratio may not be a robust measure 
of the total [Fe/H] abundance. The magnitude of these corrections is quite 
sensitive to the ionizing background, leading to some variation amongst results 
in the literature \citep{DZ03,prochaska06,peroux07,meiring07,meiring08}.
Substantial ionization corrections are, however, certainly needed in some 
sub-DLAs. The larger difference between the median $N$(\ion{H}{i}) of sub-DLAs 
and DLAs at low than high redshifts\footnote{The difference between 
the median $\log N$(\ion{H}{i}) of sub-DLAs and DLAs is 1.04~dex (0.76~dex) at 
$z<1.7$ ($z>1.7$) with the median $\log N$(\ion{H}{i}) of sub-DLAs being equal 
to 19.81 (19.94).} and the change in the shape of the ionizing background 
between $z\sim 3$ and $z\sim 1$, also raises the possibility that the typical 
ionization corrections for sub-DLAs may be higher at $z<1.7$ than $z>1.7$. 
However, there is currently no strong evidence to support this as an 
explanation for the much steeper evolution of sub-DLA metallicities. In 
general, ionization corrections in sub-DLAs are similar when calculated with a 
Haardt-Madau spectrum at both $z\sim 3$ and $z\sim 0.5$ (Milutinovic, private 
communication) and \citet{meiring07,meiring08} found corrections that are 
typically $<$ 0.2 dex for their low-redshift sub-DLAs.

Having ruled out physically related sources of bias (dust, environment, and 
ionization corrections), we consider systematics associated with the selection 
and analysis of low-redshift sub-DLAs. This is additionally motivated by our 
finding that the sub-DLAs only show significant enhancement in [Fe/H] at 
redshifts that require HST observations of the Ly$\alpha$ line. 
\citet{ellison09} have investigated possible systematic uncertainties in the 
\ion{H}{i} column density determinations of sub-DLAs from low-resolution 
spectra in comparison to high-resolution spectra. They have found that 
$N$(\ion{H}{i}) derived from low-resolution spectra tend to be over-estimated 
by typically 0.1~dex (up to 0.3~dex), but in the wrong direction to explain the 
high metallicities observed in the low-redshift sub-DLAs. 

Next, we consider the process through which the majority of our low-redshift 
sub-DLAs were identified: the \ion{Mg}{ii} equivalent width (EW) selection. 
Most of the low-redshift sub-DLAs in both this study and the literature come 
from compilations of sub-DLAs that have been obtained as a `by-product' of 
low-redshift DLA searches. These DLA surveys have used metal lines to 
pre-select candidate DLAs from ground-based spectra for HST follow-up. 
\citet{rao06} have found that there are no DLAs in their unbiased sample (i.e. 
excluding previously known 21~cm absorbers) with rest-frame 
\ion{Mg}{ii}\,$\lambda$2796 EWs lower than 0.6~\AA. Pre-selecting candidate 
DLAs based on their \ion{Mg}{ii} EWs therefore appears to be a reasonable 
selection strategy of DLAs \citep[see also][]{peroux04}. However, at lower 
\ion{H}{i} column densities, an increasing fraction of QAL systems have 
\ion{Mg}{ii}\,$\lambda$2796 EWs $< 0.6$~\AA\ \citep{rao06}. As a consequence, a 
high \ion{Mg}{ii} EW cut does not select all sub-DLAs in an unbiased way.  

In Fig.~\ref{EW-metallicity} we show the rest-frame \ion{Mg}{ii}\,$\lambda$2796 
EWs and metallicities for sub-DLAs and DLAs at $z<1.7$ of our sample and the 
sample of \citet{ellison09}. Clearly, sub-DLAs are metal-rich for their 
\ion{Mg}{ii} EWs compared with the DLAs, and this despite that the 
\ion{Mg}{ii}\,$\lambda$2796 EW distributions of sub-DLAs and DLAs at $z<1.7$ 
are statistically consistent (see their cumulative functions; KS c.l. to reject 
the null-hypothesis $= 10$\%). This offset can be explained in terms of 
kinematics. \citet{ellison06} and \citet{ellison09} showed that (low-redshift) 
sub-DLAs have higher velocity widths for a given \ion{Mg}{ii} EW than DLAs 
(parameterized by the $D$-index). Since we have already shown that $\Delta 
{\rm v}$ is correlated with metallicity in sub-DLAs 
(Fig.~\ref{velocity-metallicity}), we then expect that sub-DLAs selected on 
their high \ion{Mg}{ii} EWs will have higher metallicities. We conclude that 
the \ion{Mg}{ii} selection bias is a viable explanation for the systematically 
higher metallicities in low-redshift sub-DLAs 
\citep[but for a contrasting opinion, see][]{kulkarni07}. The ultimate test of 
this explanation would be to conduct a blind sub-DLA survey at $z<1.7$.

In summary, we conclude that in general sub-DLAs are not uniquely synonymous 
with massive galaxies and that their high metallicities observed at $z<1.7$ 
that drives an apparently steep evolution may be due to selection effects.


\section*{Acknowledgments}

We are grateful to Nikola Milutinovic for supplying results from his Cloudy 
models, and we thank Max Pettini and Jason X. Prochaska for very helpful 
exchanges. MD-Z is supported by the Swiss National Funds and SLE by an NSERC 
Discovery grant. MTM thanks the Australian Research Council for a QEII Research 
Fellowship (DP0877998).



\small{

\bsp}

\label{lastpage}

\end{document}